\documentclass{IEEEtran}
\usepackage{cite}
\usepackage{amsmath,amssymb,amsfonts}
\usepackage{algorithmic}
\usepackage{graphicx}
\usepackage{textcomp}
\usepackage [latin1]{inputenc}
\def\BibTeX{{\rm B\kern-.05em{\sc i\kern-.025em b}\kern-.08em
    T\kern-.1667em\lower.7ex\hbox{E}\kern-.125emX}}
\begin{document}
\title{A Novel, Beam-based Formalism for Active Impedance of Phased Arrays}

\author{Meiling Deng*, Ju Wu
\thanks{ M. Deng, W. Ju are with the Sichuan Jiuqiang Communications Technology Co., Ltd.  }
\thanks{This preprint is submitted to arXiv for the purpose of establishing priority. The manuscript has not been peer-reviewed or submitted to any journal. }
}

\maketitle

\begin{abstract}
The active impedance is a fundamental parameter for characterizing the behavior of large, uniform phased array antennas. However, its conventional calculation via the mutual impedance matrix (or the scattering matrix) offers limited physical intuition and can be computationally intensive. This paper presents a novel derivation of the active impedance directly from the radiated beam pattern of such arrays. This approach maps the scan-angle variation of the active impedance directly to the intrinsic angular variation of the beam, providing a more intuitive physical interpretation. The theoretical derivation is straightforward and rigorous. The validity of the proposed equation is conclusively confirmed through full-wave simulations of a prototype array. This work establishes a new and more intuitive framework for understanding, analyzing and accurately measuring the scan-dependent variations in phased arrays, which is one of the main challenges in modern phased array designs. Consequently, this novel formalism is expected to expedite and simplify the overall design and optimization process for next-generation, large-scale uniform phased arrays.
\end{abstract}


\section{Introduction}
The deployment of large-scale, uniform phased arrays has become the transformative technology across critical domains: high-throughput satellite communications \cite{sate}, advanced radar systems \cite{radar1} \cite{radar2}, next-generation radio telescopes \cite{chime} and massive MIMO for 5G/6G wireless communications \cite{5gb1} \cite{5gb2} \cite{5gb3} \cite{5gb4}.  Their supremacy stems from the ability to electronically shape and steer high-gain beams with unparalleled agility. Furthermore, supported by sophisticated digital backend processors, these arrays can generate multiple independent beams simultaneously, enabling capabilities such as real-time multi-target tracking and concurrent multi-sector communication. This technology is fundamentally redefining the standards for spectral efficiency, connectivity density, spatial resolution and system gain across the electromagnetic spectrum.

Harnessing this potential, however, hinges on mastering a fundamental physical challenge: the intricate electromagnetic mutual coupling between antenna elements within the array \cite{mc}. This coupling dictates that the impedance "seen" by each radiating element -- termed as active impedance -- is not a fixed property as that of a single antenna but can dynamically vary with the array's excitation, particularly the scan angle of the synthesized beam. Consequently, the active impedance serves as the critical bridge linking the macroscopic array performance (gain, scanning blindness, bandwidth) to the microscopic behavior of how individual elements match to the arrays' feeding network under different excitation states.  Accurate knowledge and control of this parameter are therefore paramount for achieving robust and optimal array operation.

Conventionally, the active impedance is determined indirectly through the framework of network theory. It is calculated from the complete mutual impedance matrix (Z) or equivalently derived from the full-wave simulated or measured scattering matrix (S) of the array. While this approach is natural at first sight and forms the established backbone of contemporary design workflows, it suffers from two inherent and significant drawbacks. 

First, it lacks physical intuition: the mutual impedance matrix acts as an abstract "black-box" network representation. From a design perspective, the impedance matrix formalism presents a fundamental disconnect between cause and effect. While a designer's primary intent can be to control the array's gain at specific directions, any physical adjustment to the array (e.g., modifying an element's geometry or element spacing) manifests as opaque, non-intuitive alterations across numerous entries of the mutual impedance matrix. There is no direct or clear physical insight linking these scattered matrix perturbations to the resultant change of the array's macroscopic performance at a specific, targeted direction of interest. This abstraction forces designers to rely on random iterative simulations rather than guided intuition. Second, the conventional approach can be computationally and practically inefficient. In the simulation phase, obtaining the full S- or Z-parameter matrix requires high-fidelity modelling of each individual antenna port. However, the far-field radiation pattern -- the ultimate performance metric -- is primarily governed by the overall current distribution on the array, which does not necessitate the same level of port-level precision. This can lead to unnecessary computational overhead, especially for large phased arrays with fine structure at antenna ports. In the experimental phase, direct measurement of the active impedance for various scan states requires a complete and complex feeding network capable of applying all required excitation vectors, making the characterization process prohibitively time-consuming and costly for large arrays. Consequently, this indirect paradigm makes the entire workflow -- from analysis and optimization to diagnostic testing -- cumbersome for rapid development cycles of large-scale phased arrays.

To overcome these limitations, recent studies have sought to connect impedance to radiation characteristics \cite{za1} \cite{za2}, yet these formulations are either not the direct design target, or the formulation is not practical enough compared to the conventional method. Building upon insights from these two studies, we present another novel theoretical formalism that establishes a closed-form expression for the active impedance directly from the radiated beam field of the array. This approach bypasses the intermediary network parameters altogether. The proposed formulation elegantly maps the scan-angle variation of the active impedance directly to the intrinsic angular variation of array's beam that is easy and straightforward to simulate and measure, even without the need to include the array's matching load. Beyond its theoretical elegance which introduces new insights on how large uniform phased array works, this beam-centric perspective points to a much more direct and efficient measurement methodologies without the need for complete and complex feeding network, potentially streamlining the characterization and design cycles for next-generation phased array systems.

This paper is organized as follows. In Section 2 we present the core theoretical contribution: the novel derivation of the active impedance starting from the fundamental principles. A closed-form formula is developed, explicitly linking the active impedance to the observable beam pattern. In Section 3 we present the simulation setup and its comprehensive results from full-wave electromagnetic analysis of a representative phased array. Based on the simulation data, the active impedance obtained from conventional method is compared with the predictions of the derived formula across various scan angles, demonstrating conclusive agreement within simulation uncertainty and thereby validating the proposed theory. Finally, Section 4 concludes the paper by summarizing the key findings, reiterating the advantages of the beam-based approach, and discussing its implications for future array design and characterization.

\section{Theoretical Derivation}
\label{sec:2nd}
\begin{figure}[h!]
\centering
{
  \includegraphics[width=0.45\textwidth]{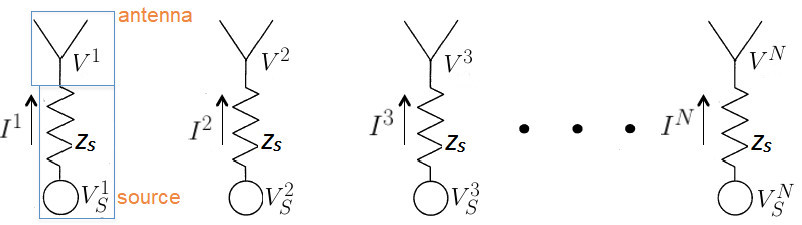}
}
\caption{the Thevevin-equivalent circuit of a phased array and its feeding network system. The array consists of N elements. At each port, the feeding network is equalized as a voltage source in series with a source input impedance $z_S$.}
\label{circuit}
\end{figure}

To derive the active impedance, we assume the phased array is working in the transmitting mode throughout this paper due to reciprocity. Fig. \ref{circuit} shows the Thevevin-equivalent circuit diagram of a phased array system with its feeding network in the transmitting mode. In this diagram, each antenna element is in series with a voltage source and a source input impedance $z_S$.\\

According to circuit theory, :
\begin{equation}
\label{eq1}
V^S=z_SI+V
\end{equation}
\begin{equation}
\label{eq2}
V=\mathbf{Z}I
\end{equation}
, where $V^S$ is a vector with its entry $V^S_n$ being the source voltage exciting the $n^{th}$ antenna element, $I$ is a vector with its entry $I_n$ being the port current of the $n^{th}$ antenna element, $V$ is a vector with its entry $V_n$ being the port voltage of the $n^{th}$ antenna element,  and $\mathbf{Z}$ is the impedance matrix of the phased array.

From Eqn. \ref{eq1} and \ref{eq2},
\begin{equation}
\label{ivs}
I=(z_S+\mathbf{Z})^{-1} V^S
\end{equation}
\begin{equation}
\label{vvs}
V=\mathbf{Z} (z_S+\mathbf{Z})^{-1} V^S
\end{equation}

Now consider the $m^{th}$ antenna's active element pattern $E^S_m$, which is defined as the beam of the phased array when only the $m^{th}$ antenna is active while all others are passive so that source voltage satisfies $V^S_n=\delta_{mn}$. Based on linearity of Maxwell Equation,
\begin{equation}
\label{gve0}
E^S_m(\theta,\phi) =\sum_n V_n E^V_n(\theta,\phi)
\end{equation}
, where $E^V_n(\theta,\phi)$ is short-circuit pattern defined as the beam of the phased array with antenna port voltage $V_k=\delta_{kn}$. Due to the translation symmetry of uniform phased arrays,
 \begin{equation}
 \label{gve1}
 E^V_n(\theta,\phi) = E^V_0(\theta,\phi) e^{i \vec{k}(\theta, \phi) \cdot \vec{r}_{n}}
 \end{equation}
 , where $\vec{k}(\theta, \phi)$ is the free-space propagation vector, $\vec{r}_{n}$ is the position vector of the $n^{th}$ element in the array. \\
 
Combine Eqn. \ref{vvs}, Eqn. \ref{gve0} and Eqn. \ref{gve1}, $E^S_m$ can be rewritten as 
 \begin{equation}
\label{gve}
\begin{split}
E^S_m(\theta,\phi)  &=E^V_0(\theta,\phi) \sum_{n} V_n e^{i \vec{k}(\theta, \phi) \cdot \vec{r}_{n}}\\
 &=E^V_0(\theta,\phi)  \sum_n \{\sum_j  [\mathbf{Z} (z_S+\mathbf{Z})^{-1}]_{nj} V^S_j\} e^{i\vec{k}(\theta, \phi) \cdot \vec{r}_{n}} \\
 &= E^V_0(\theta,\phi)  \sum_{n} [\mathbf{Z} (z_S+\mathbf{Z})^{-1} ]_{mn} e^{i\vec{k}(\theta, \phi) \cdot \vec{r}_{n}} \\
 &= E^V_0(\theta,\phi) V_m(when  \quad V^S_n=e^{i\vec{k}(\theta, \phi) \cdot \vec{r}_{n}})
\end{split}
\end{equation}

Eqn. \ref{gve0} is based on linearity of port voltage. Similarly, active element pattern $E^s_m$ can be derived based on linearity of port current as
\begin{equation} 
\label{gie0}
E^S_m(\theta,\phi)  =\sum_n I_n E^I_n(\theta,\phi),
\end{equation}
 \begin{equation}
 \label{gie1}
 E^I_n(\theta,\phi) = E^I_0(\theta,\phi) e^{i \vec{k}(\theta, \phi) \cdot \vec{r}_{n}}
 \end{equation}
, where $E^I_n(\theta,\phi)$ is open-circuit pattern defined as the beam of the phased array with antenna port current $I_k=\delta_{kn}$. \\

Combining Eqn. \ref{ivs}, Eqn. \ref{gie0} and Eqn. \ref{gie1}, $E^S_m$ can also be rewritten as
 \begin{equation}
\label{gie}
\begin{split}
E^S_m(\theta,\phi)  &=E^I_0(\theta,\phi) \sum_{n} I_n e^{i \vec{k}(\theta, \phi) \cdot \vec{r}_{n}}\\
 &=E^I_0(\theta,\phi)  \sum_n \{\sum_j  [(z_S+\mathbf{Z})^{-1}]_{nj} V^S_j\} e^{i\vec{k}(\theta, \phi) \cdot \vec{r}_{n}} \\
 &= E^I_0(\theta,\phi)  \sum_{n} [(z_S+\mathbf{Z})^{-1} ]_{mn} e^{i\vec{k}(\theta, \phi) \cdot \vec{r}_{n}} \\
 &= E^I_0(\theta,\phi) I_m(when  \quad V^S_n=e^{i\vec{k}(\theta, \phi) \cdot \vec{r}_{n}})
\end{split}
\end{equation}

Eqn. \ref{gve} and \ref{gie} together lead to 
\begin{equation}
\label{za}
\begin{split}
z_a(\theta,\phi) 
&\equiv \frac{V_m(when \quad V^S_n=e^{-i\vec{k}(\theta, \phi) \cdot \vec{r}_{n}})}{I_m(when \quad V^S_n=e^{-i\vec{k}(\theta, \phi) \cdot \vec{r}_{n}})} \\
&= \frac{V_m(when \quad V^S_n=e^{i\vec{k}(\pi-\theta, \phi-\pi) \cdot \vec{r}_{n}})}{I_m(when \quad V^S_n=e^{i\vec{k}(\pi-\theta, \phi-\pi) \cdot \vec{r}_{n}})} \\
&=  \frac{E^I_0(\pi-\theta, \phi-\pi) }{E^V_0(\pi-\theta, \phi-\pi)}  \\
&= \frac{E^I_0(\theta, \phi-\pi) }{E^V_0(\theta, \phi-\pi)} (for\ planar\ or\ linear\ array) \\
&= \frac{E^I_0(\theta, \pi) }{E^V_0(\theta, \pi)} (for\ symmetric\ array) \\
&= \frac{E^I_m(\theta, \pi) }{E^V_m(\theta, \pi)}, \forall m
\end{split}
\end{equation}
, where $z_a(\theta, \phi)$ is the active impedance when the array is excited with $V^S_n=e^{-i\vec{k}(\theta, \phi) \cdot \vec{r}_{n}}$ so that the beam is synthesized to the direction $\vec{k}(\theta, \phi)$. \\

Eqn. \ref{za} is our final formalism, which shows how to derive active impedance of large, uniform array from only two sets of beam data, one open-circuit pattern when all other elements in the array are open-loaded while the other short-circuit pattern when all other elements in the array are short-loaded. \\

Based on Eqn. \ref{za}, its also very straightforward to derive another important equation of embedded element pattern from the short-circuit pattern and open-circuit pattern. To see this how it works, let's re-write Eqn. \ref{gve} with the scattering matrix approach rather than the impedance matrix approach.
\begin{equation}
\label{eep}
\begin{split}
E^S_m(\theta,\phi)  &=E^V_0(\theta,\phi) \sum_{n} V_n e^{i \vec{k}(\theta, \phi) \cdot \vec{r}_{n}}\\
& =E^V_0(\theta,\phi) \sum_{n} (V^+_n + V^-_n) e^{i \vec{k}(\theta, \phi) \cdot \vec{r}_{n}}\\
& =E^V_0(\theta,\phi) (e^{i \vec{k}(\theta, \phi) \cdot \vec{r}_{m}}+\sum_{n}S_{nm}e^{i \vec{k}(\theta, \phi) \cdot \vec{r}_{n}})\\
& =E^V_0(\theta,\phi)e^{i \vec{k}(\theta, \phi) \cdot \vec{r}_{m}} (1+\sum_{n}S_{nm}e^{i \vec{k}(\theta, \phi) \cdot (\vec{r}_n-\vec{r}_m)})\\
& =E^V_0(\theta,\phi)e^{i \vec{k}(\theta, \phi) \cdot \vec{r}_{m}} (1+\Gamma_{a})\\
& = E^V_0(\theta,\phi)e^{i \vec{k}(\theta, \phi) \cdot \vec{r}_{m}} (1+\frac{z_a-z_s}{z_a+z_s})\\
& =E^V_m \frac{2z_a}{z_a+z_s}\\
& =2 \frac{E^I_m E^V_m}{E^I_m+z_sE^V_m}
\end{split}
\end{equation}

\section{Simulation Data}
\label{sec:3rd}
To verify the validity of Eqn. \ref{za} and Eqn. \ref{eep}, a 15-element linear array is simulated with HFSS. Each array element is a planar dipole antenna that is oriented parallel to the y-axis, and the array elements are uniformly spaced along the x-axis. More detailed information of this project is listed in table \ref{tab1}. \\
\begin{table}
\caption{simulation setup of the prototype array}
\label{table}
\setlength{\tabcolsep}{3pt}
\begin{tabular}{|p{50pt}|p{90pt}|p{50pt}|}
\hline
item    &  property   &  value [units] \\
\hline
dipole element   & arm length & 150 [mm] \\
dipole element   & arm width & 5 [mm] \\
dipole element   & feeding gap & 2 [mm] \\
array geometry & central element position & [0, 0, 0] [mm] \\
array geometry  & element spacing & 150 [mm] \\
sim. setup &frequency &500 [MHz]\\
sim. setup &convergence criteria &0.02\\
\hline
\end{tabular}
\label{tab1}
\end{table}

In total, we have run 3 simulations with 3 different source impedance setups: 1, all elements are excited with lumped port with port impedance 50 Ohm to get he embedded element pattern at 50 Ohm load and the S-parameters of the array for the calculation of active impedance; 2, all lumped ports except for the central one, are changed to a lump impedance of resistance $R=1000,000$ to get the open-circuit pattern $E^I_0$; 3, all lumped ports except for the central one, are changed to a PEC which connects the two arms of each dipole antenna to get the short-circuit pattern $E^V_0$. Note, since each element is aligned along y-axis, we choose to use the y-component of the farfield beam for the verification of Eqn. \ref{za} and Eqn. \ref{eep} as both of them hold true for any component of the beam data.\\

\subsection{active impedance derivation verification}
\begin{figure}[h!]
\centering
{
  \includegraphics[width=0.5\textwidth]{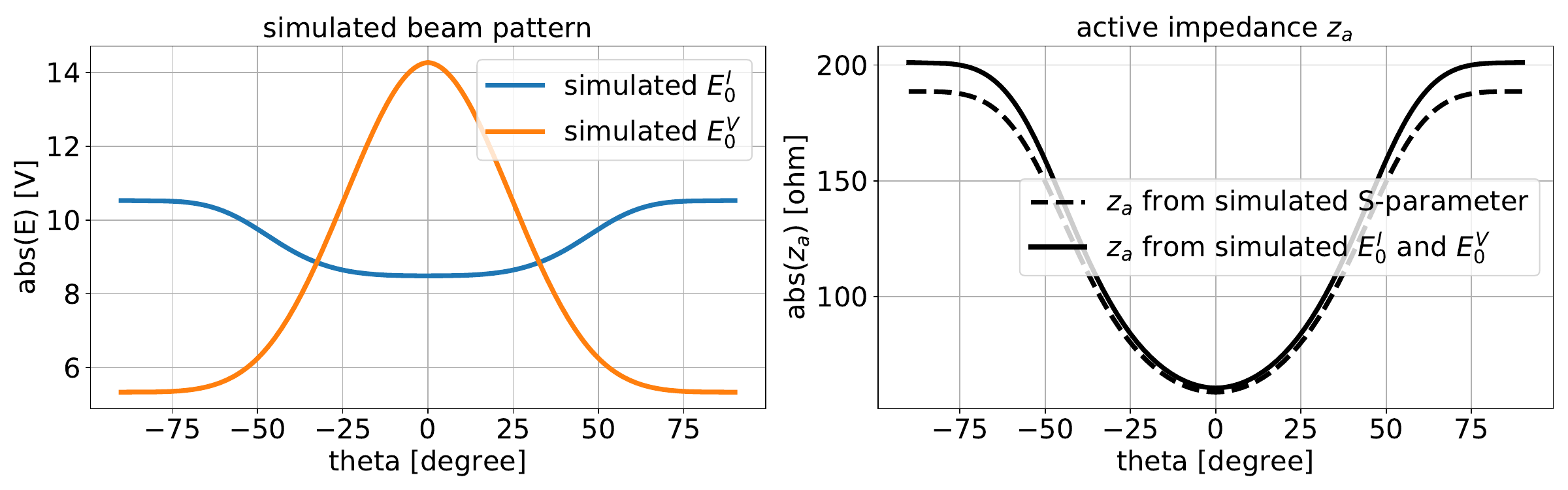}
}
\caption{The active impedance $z_a$ derived from conventional S-parameters and that from the beam with short/open load are compared along the $\phi=0$ slice at various $\theta$ scanning angle. The left panel shows the beam data itself that is used for the derivation of the active impedance. The right panel shows the magnitude comparison between the the active impedance $z_a$ from conventional S-parameters and $z_a$ from beam data. }
\label{zaphi0}
\end{figure}

In Fig. \ref{zaphi0}, the active impedance derived from the conventional S-parameters and that derived from open-circuit beam and short-circuit beam is compared at various $\theta$ along the slice $\phi = 0$. It shows how the pattern of active impedance in the right panel can be reproduced from the other two distinctive patterns of $E^I$ and $E^V$ in the left panel of Fig. \ref{zaphi0}. The observed larger disagreement as angle moves towards the horizon (ie, $\theta=90$) could be due to the following two reasons: first, as the relative error is still on the scale of of $\sim 5\%$ across the whole range, so the absolute disagreement becomes larger as the absolute value of active impedance becomes larger when the angle approaching the horizon of the array; second, the calculation of $z_a$ from multiple S-matrix entries involves complex error propagation which varies with angle. Given that the simulation convergence is set to $2\%$, the disagreement on the scale of $5\%$ is acceptable considering this complicated error propagation from various S-matrix entries. Moreover, the nearly perfect match shown in Fig. \ref{eepphi0} which validates Eqn. \ref{eep} clearly excludes the scenario that the larger disagreement at horizon is from theoretical derivation itself as Eqn. \ref{eep} is purely based on Eqn. \ref{za}. \\

\begin{figure}[h!]
\centering
{
  \includegraphics[width=0.5\textwidth]{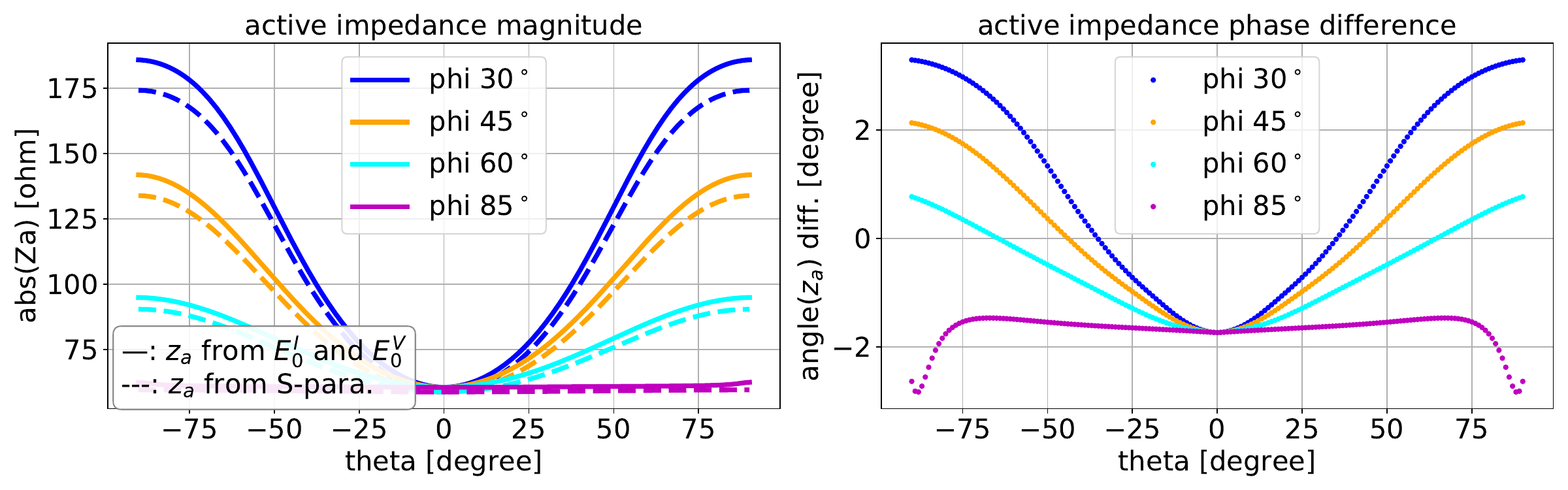}
}
\caption{The active impedance $z_a$ derived from conventional S-parameters and that from the beam with short/open load are compared along multiple $\phi$ slices. The left panel shows the magnitude comparison. The right panel shows the phase difference between these two methods, which are mostly within 2 degrees.}
\label{zaphis}
\end{figure}

Fig. \ref{zaphis} shows both the magnitude and phase comparison of the active impedance derived from the two methods at other angles. The magnitude comparison at other $\phi$ slices are similar to that at $\phi=0$. It's of particular interest to see that when $\phi \rightarrow 90$, the active impedance from conventional S-parameters should be constant across $\theta$ range due to its mathematical definition, while the active impedance derived from $\frac{E^I}{E^V}$ successfully replicates this constant line. As for the phase comparison, it could be seen that $z_a$'s phase difference between this two methods are mostly within 2 degrees, which is within the simulation uncertainty.

\subsection{embedded element pattern derivation verification}
\begin{figure}[h!]
\centering
{
  \includegraphics[width=0.5\textwidth]{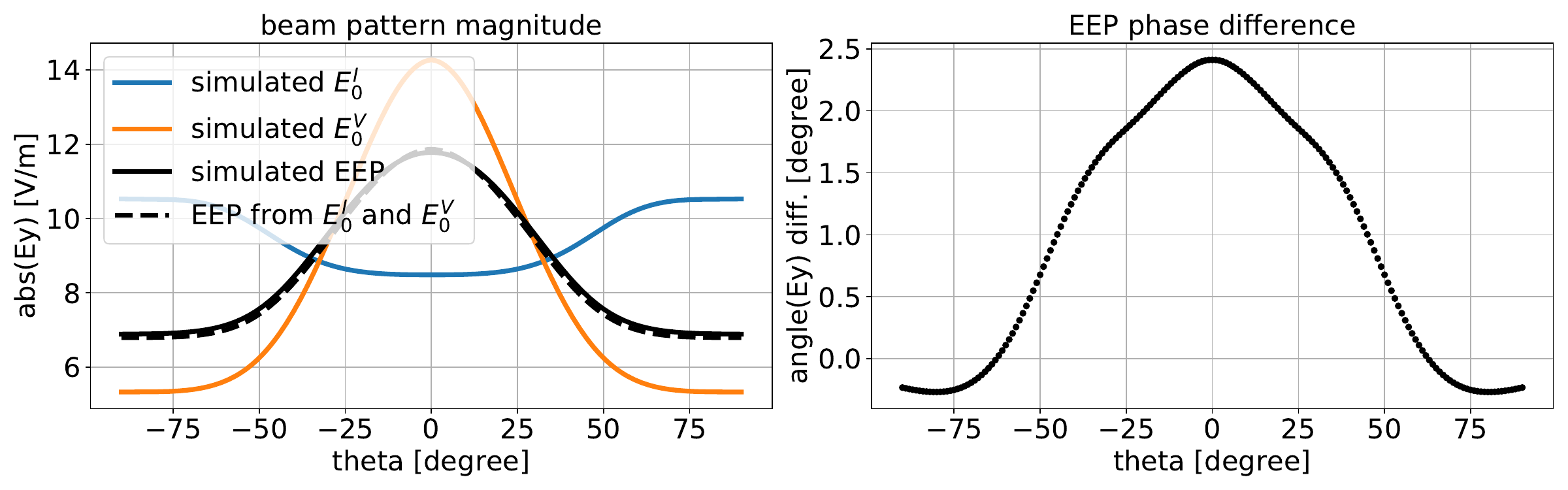}
}
\caption{The directly simulated embedded element pattern and the beam-derived embedded element pattern are compared along the $\phi=0$ slice at various $\theta$ angle. In the left panel, the magnitude of these two embedded element pattern together with the directly simulated $E^I$ and $E^V$ are plotted together. The right panel shows the phase difference between the directly simulated and beam-derived embedded element pattern. }
\label{eepphi0}
\end{figure}

\begin{figure}[h!]
\centering
{
  \includegraphics[width=0.5\textwidth]{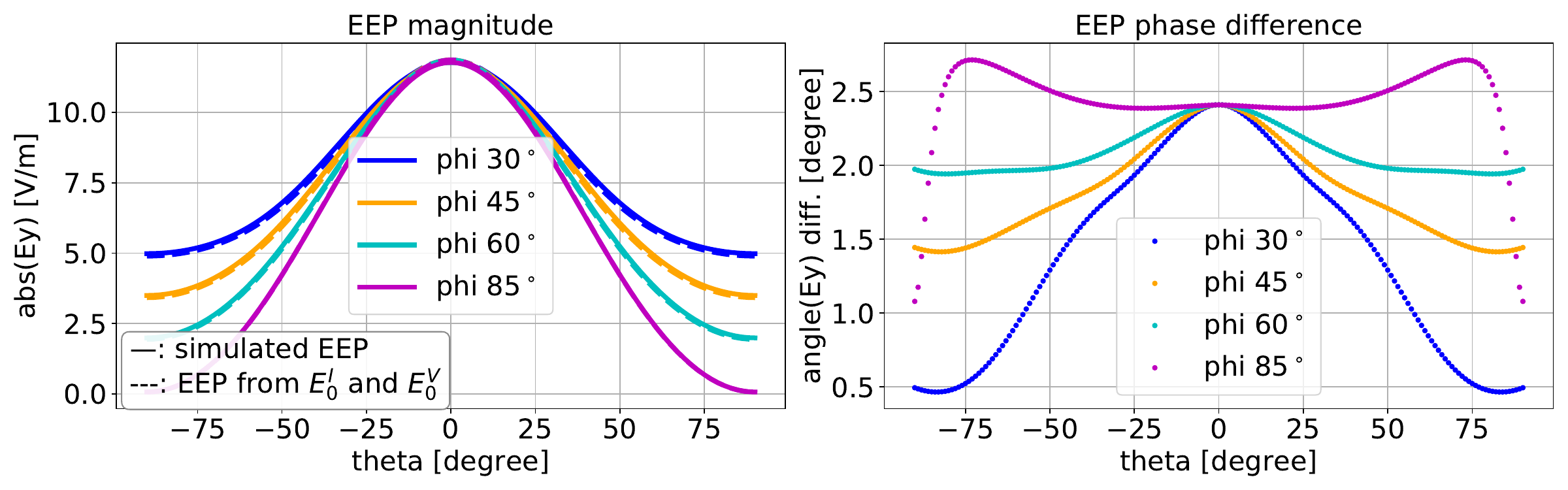}
}
\caption{The directly simulated embedded element pattern and the beam-derived embedded element pattern are compared along multiple $\phi$ slices at various $\theta$ angle. In the left panel, the magnitude of these two embedded element pattern together with the directly simulated $E^I$ and $E^V$ are plotted together. The right panel shows the phase difference between the directly simulated and beam-derived embedded element pattern.}
\label{eepphis}
\end{figure}

To check the validity of Eqn. \ref{eep}, the simulated and derived embedded element pattern at slice $\phi=0$ are plotted in Fig. \ref{eepphi0}, which shows nearly a perfect match between these two patterns. The nearly perfect match at other $\phi$ slices, as shown in Fig. \ref{eepphis}, further strongly proves the theoretical derivation of embedded element pattern from the open-circuit pattern and the short-circuit pattern. 

\section{Conclusion}
This paper has presented a novel, beam-based formalism for characterizing the active impedance and embedded element pattern in large uniform phased arrays. A closed-form expression was rigorously derived, directly linking the scan-dependent impedance to the intrinsic angular variation of array beams. The validity of the proposed formulation were conclusively demonstrated through full-wave electromagnetic simulations of a 15-element phased array. This work shifts the paradigm for understanding mutual coupling from a network-centric (ie, the impedance or scattering matrix approach) to a radiation-centric perspective. This fundamental shift provides deeper physical insight into phased array theory, strengthens intuitive guidance during simulation and design, and points the way toward more efficient and accurate experimental characterization. 





\end{document}